\documentclass[12pt]{article}
\usepackage{cite}
\usepackage{feynmp-auto}
\usepackage{slashed}
\usepackage[pdftex]{hyperref}
\usepackage{amsmath, amsthm, amssymb,slashed,mathrsfs}
\parskip=\baselineskip

\font\teneurm=eurm10 \font\seveneurm=eurm7 \font\fiveeurm=eurm5
\newfam\eurmfam
\textfont\eurmfam=\teneurm \scriptfont\eurmfam=\seveneurm
\scriptscriptfont\eurmfam=\fiveeurm

 \font\teneusm=eusm10 \font\seveneusm=eusm7 \font\fiveeusm=eusm5
\newfam\eusmfam
\textfont\eusmfam=\teneusm \scriptfont\eusmfam=\seveneusm
\scriptscriptfont\eusmfam=\fiveeusm

\font\tencmmib=cmmib10 \skewchar\tencmmib='177
\font\sevencmmib=cmmib7 \skewchar\sevencmmib='177
\font\fivecmmib=cmmib5 \skewchar\fivecmmib='177
\newfam\cmmibfam
\textfont\cmmibfam=\tencmmib \scriptfont\cmmibfam=\sevencmmib
\scriptscriptfont\cmmibfam=\fivecmmib

\parskip=\baselineskip

\def\e{\mbox{\boldmath$\displaystyle\mathbf{\epsilon}$}}
\def\p{\mbox{\boldmath$\displaystyle\mathbf{p}$}}

\def\th{\mbox{\boldmath$\displaystyle\mathbf{\theta}$}}

\def\bv{\mbox{\boldmath$\displaystyle\mathbf{\varphi}$}}

\def\0{\mbox{\boldmath$\displaystyle\mathbf{0}$}}
\def\s{\mbox{\boldmath$\displaystyle\mathbf{\sigma}$}}
\def\J{\mbox{\boldmath$\displaystyle\mathbf{J}$}}

\def\x{\mbox{\boldmath$\displaystyle\mathbf{x}$}}

\def\y{\mbox{\boldmath$\displaystyle\mathbf{y}$}}

\newcommand{\dual}[1]{\overset{{}^{{}^{\boldsymbol{\neg}}}}{\smash[t]{#1}}} 
\def\df{\mbox{$\displaystyle\gdual{\mathfrak{f}}$}}
\newcommand{\gdual}[1]{\overset{\:{}^{{}^{\boldsymbol{\neg}}}}{\smash[t]{#1}}}

\begin{document}
\unitlength=1mm
\vskip 1.5in
\begin{center}
{\bf\Large{Spin-half mass dimension one fermions and their higher-spin generalizations}}\vskip 0.5cm

{Cheng-Yang Lee} 

{\small{\textit{Center for Theoretical Physics, College of Physical Science and Technology,\\
Sichuan University, Chengdu, 610064, China\\
Email: cylee@scu.edu.cn}\\
}} \vskip 0.1in

\end{center}
\vskip 0.3in

\begin{abstract}
A self-contained review on spin-half mass dimension one fermions and their higher-spin generalizations is presented. Starting from the two-component left-handed Weyl spinors, the Dirac spinors and Elko (eigenspinors of the charge conjugation operator) are constructed. After elaborating on their similarities and differences, we generalize the spin-half Elko to higher-spin. The field operators constructed from Elko and their higher-spin generalizations are shown to be of mass dimension one with positive-definite free Hamiltonians. The physical significance of higher-spin mass dimension one particles and further extensions in the context of Lounesto classification are discussed.
\end{abstract}

\section{Introduction}

In 1928, Dirac wrote down an equation that revolutionized our understanding of the universe~\cite{Dirac:1928hu}. Ninety years later, it remains to be one of the most beautiful equations in science. Enamoured by its rich mathematical structure and empirical successes, an unspoken consensus has quietly but surely permeated through the scientific community - a massive spin-half fermion must be described by the Dirac equation.

In 2005, Ahluwalia and Grumiller made a fundamental theoretical discovery - a massive spin-half fermion of mass dimension one~\cite{Ahluwalia:2004ab,Ahluwalia:2004sz}. This was an unexpected result contrary to the longstanding consensus with important implications for quantum field theory (QFT) and physics beyond the Standard Model (SM). These fermions have two surprising properties - They satisfy the Klein-Gordon but not the Dirac equation and are of mass dimension one instead of three-half. Therefore, they have renormalizable quartic self-interactions, making them potential dark matter candidates.

What underlies this construct was a structure already known to mathematicians studying spinors and Clifford algebra~\cite{Lounesto:2001zz}. A systematic classification shows that the Dirac and Weyl spinors are not the only possible spin-half representations of the Lorentz group. In fact, there exists two other classes of spinors known as the flag-pole and flag-dipole spinors. Elko, the eigenspinor of the charge-conjugation operator associated with mass dimension one fermions, was shown to be a flag-pole spinor~\cite{daRocha:2005ti}.

The works of Ahluwalia and Grumiller have received increasing attentions in various areas ranging from cosmology~\cite{Boehmer:2006qq,Boehmer:2007dh,Boehmer:2008rz,Boehmer:2007ut,Boehmer:2008ah,Boehmer:2009aw,Boehmer:2010tv,Shankaranarayanan:2009sz,Shankaranarayanan:2010st,Boehmer:2010ma,Gredat:2008qf,Wei:2010ad,Basak:2012sn,daSilva:2014kfa}, quantum field theory~\cite{Ahluwalia:2008xi,Ahluwalia:2009rh,Lee:2012td,Lee:2014opa,Lee:2015sqj,Ahluwalia:2016rwl}, particle phenomenologies~\cite{Dias:2010aa,Alves:2014kta,Alves:2014qua,Agarwal:2014oaa,Alves:2017joy} and mathematics~\cite{daRocha:2005ti,daRocha:2008we,daRocha:2007pz,HoffdaSilva:2009is,daRocha:2009gb,daRocha:2011yr,daRocha:2011xb,Bernardini:2012sc,daRocha:2013qhu,Ablamowicz:2014rpa,Bonora:2014dfa,Bueno_Rogerio_2017}. The works accomplished so far strongly suggest that the relationships between spinors and QFT have an even deeper and richer structure than previously expected.

In this paper, we present a self-contained review on the construct of spin-half Elko, mass dimension one fermions and their higher-spin generalizations. Our construct begins by constructing and comparing the Dirac spinors and Elko. This naturally leads to a new set of duals for Elko that is different from its Dirac counterpart. By construction, we show that Elko do not satisfy the Dirac equation and that their spin-sums are not Lorentz-covariant. The non-covariance can be removed by introducing a new dual with an infinitesimal deformation~\cite{Ahluwalia:2016rwl}.

The spin-half construct has a natural higher-spin generalizations~\cite{Lee:2012td}. The resulting higher-spin particles have mass dimension one and are physically well-defined with positive-definite free Hamiltonians. The bosonic fields are however, non-local. In this paper, the non-locality is resolved by a careful choice of coefficients.

The paper is organized as follows. In sec.~\ref{de}, we review the construction of Dirac spinors and Elko. Particular emphasis is placed on their similarities and differences. In sec.~\ref{mdof}, the spin-half mass dimension one field and its adjoint are constructed. The kinematics and locality structure are discussed in detail.
In sec.~\ref{hs}, we review the results in ref.~\cite{Lee:2012td} where the spin-half construct was generalized to higher spin.  This work established the fact that the spin-half construct is only a special case. There exists an infinite tower of higher-spin particles of mass dimension one with well-defined physical properties. 

\section{The Dirac spinors and Elko} \label{de}

Our story begins with the two-component left-handed Weyl spinors of the $(\frac{1}{2},0)$ representation of the Lorentz group in the helicity basis. Let $\chi(\e,\sigma)$ be the rest spinor where $\e$ denotes the direction of the spin-projection and $\sigma=\pm\frac{1}{2}$ being the eigenvalues of $\mathbf{J}\cdot\e$. In the chosen representation, we have $\mathbf{J}=\frac{1}{2}\s$ so that
\begin{equation}
\left(\frac{1}{2}\s\cdot\e\right)\chi(\e,\sigma)=\sigma\chi(\e,\sigma). \label{eq:hel}
\end{equation}
The positive and negative eigenvalues have the physical interpretation that the spins are parallel and anti-parallel with respect to $\e$. We choose to work in the spherical coordinate with 
\begin{equation}
\e=(\sin\theta\cos\phi,\sin\theta\sin\phi,\cos\theta)
\end{equation}
where $0\leq\theta\leq\pi$, $0\leq\phi<2\pi$ are the polar and azimuthal angles respectively. The solutions to eq.~(\ref{eq:hel}), with the appropriate normalization are given by
\begin{equation}
{\chi(\e,\textstyle{\frac{1}{2}})}=\sqrt{\frac{m}{2}}\left[\begin{array}{c}\cos(\theta/2)e^{-i\phi/2} \\
\sin(\theta/2)e^{i\phi/2}\end{array}\right],\quad
{\chi(\e,\textstyle{-\frac{1}{2}})}=\sqrt{\frac{m}{2}}
\left[\begin{array}{c}
-\sin(\theta/2)e^{-i\phi/2} \\
\cos(\theta/2)e^{i\phi/2}\end{array}\right] \label{eq:phi}
\end{equation}
with $m$ being the mass of the particle.

The boost and rotation of $\chi(\e,\sigma)$ are given by
\begin{eqnarray}
\chi(\p,\sigma)&=&\exp(i\mathbf{K}\cdot\bv)\chi(\e,\sigma)\nonumber\\
&=&\exp\left(-\frac{1}{2}\s\cdot\bv\right)\chi(\e,\sigma) \label{eq:lb}
\end{eqnarray}
and
\begin{eqnarray}
\chi(\mathbf{R}\e,\sigma)&=&\exp(i\mathbf{J}\cdot\th)\chi(\e,\sigma) \nonumber\\
&=&\exp\left(\frac{i}{2}\s\cdot\th\right)\chi(\e,\sigma).
\end{eqnarray}
In eq.~(\ref{eq:lb}), the boost generator is $\mathbf{K}=i\s/2$ and the rapidity parameter is defined as $\bv=\hat{\p}\,\varphi$, $\cosh\varphi=E/m$, $\sinh\varphi=|\p|/m$. Since we are working in the helicity basis, the direction of boost can either be parallel $\hat{\p}=\e$ or anti-parallel $\hat{\p}=-\e$ with respect to the spin. In eq.~(\ref{eq:lb}), we have chosen $\hat{\p}=\e$.

The right-handed Weyl spinors of the $(0,\frac{1}{2})$ representation can also be constructed in the same manner as above. They have the same rotation transformation as $\chi(\e,\sigma)$ but the boost is given by\footnote{Here, $\rho(\e,\sigma)$ are eigenspinors of $\J\cdot\e$ so their solutions are also given by eq.~(\ref{eq:phi}). The only difference from $\chi(\e,\sigma)$ is the boost transformation.}
\begin{equation}
\rho(\p,\sigma)=\exp\left(\frac{1}{2}\s\cdot\bv\right)\rho(\e,\sigma). \label{eq:rb}
\end{equation}
As we will show below, for the construction of Dirac spinors and Elko, the relationships between the left- and right-handed spinors are very important.

\subsection{Dirac spinors} \label{Dirac}

We will now construct the Dirac spinors and derive the field equation.  From eqs.~(\ref{eq:lb}) and (\ref{eq:rb}), we observe that under parity $\p\rightarrow-\p$,\footnote{In spherical coordinate, this means $\theta\rightarrow\pi-\theta$, $\phi\rightarrow\phi\pm\pi$ where the top and bottom sign applies when the $y$-comonent of $\p$ is positive or negative respectively.} the spinor $\rho(-\p,\sigma)$ transforms as a left-handed spinor
\begin{equation}
\rho(-\p,\sigma)=\exp\left(-\frac{1}{2}\s\cdot\bv\right)\rho(\e,\sigma).\label{eq:lb1}
\end{equation}
Therefore, using eqs.~(\ref{eq:rb}) and (\ref{eq:lb1}), we can construct a four-component spinor that transforms under the $(\frac{1}{2},0)\oplus(0,\frac{1}{2})$ representation of the form\footnote{Our choice for constructing $\psi(\p,\sigma)$ is a matter of convention, it is also possible to use $\chi(\e,\sigma)$. In the next section, we will construct Elko using $\chi(\e,\sigma)$.}
\begin{equation}
\psi(\p,\sigma)=
\left[\begin{array}{c}
\rho(\p,\sigma) \\
\alpha\rho(-\p,\sigma)\end{array}\right]\label{eq:ds}
\end{equation}
where $\alpha$ is a phase to be determined. When $\alpha=\pm1$, $\psi(\p,\sigma)$ becomes the Dirac spinor and satisfies the Dirac equation. To see this, we rewrite eq.~(\ref{eq:ds}) as
\begin{eqnarray}
\left[\begin{array}{c}
\rho(\p,\sigma) \\
\alpha\rho(-\p,\sigma)\end{array}\right]&=&\alpha
\left[\begin{matrix}
\exp\left(\frac{1}{2}\s\cdot\bv\right) & O \\
O & \exp\left(-\frac{1}{2}\s\cdot\bv\right)
\end{matrix}\right]
\left[\begin{matrix}
\alpha\rho(\e,\sigma) \\
\rho(\e,\sigma)
\end{matrix}\right]\nonumber\\
&=&\alpha\left[\begin{matrix}
\exp\left(\frac{1}{2}\s\cdot\bv\right) & O \\
O & \exp\left(-\frac{1}{2}\s\cdot\bv\right)
\end{matrix}\right]
\left[\begin{matrix}
O & I \\
I & O
\end{matrix}\right]
\left[\begin{matrix}
\rho(\e,\sigma) \\
\alpha\rho(\e,\sigma)
\end{matrix}\right]\nonumber\\
&=&\alpha\left[\begin{matrix}
O & \exp\left(\s\cdot\bv\right)\\
 \exp\left(-\s\cdot\bv\right) & O
\end{matrix}\right]
\left[\begin{matrix}
\rho(\p,\sigma) \\
\alpha\rho(-\p,\sigma)
\end{matrix}\right].
\end{eqnarray}
A straightforward calculation yields
\begin{equation}
\left[\begin{matrix}
O & \exp\left(\s\cdot\bv\right) \\
\exp\left(-\s\cdot\bv\right) & O
\end{matrix}\right] =\frac{1}{m}\gamma^{\mu}p_{\mu}
\end{equation}
where
\begin{equation}
\gamma^{0}=\left(\begin{matrix} 
O & I \\
I & O\end{matrix}\right),\quad
\gamma^{i}=\left(\begin{matrix}
O & -\sigma^{i} \\
\sigma^{i} & O\end{matrix}\right).
\end{equation}
Therefore, $\psi(\p,\sigma)$ satisfies the Dirac equation when $\alpha=\pm1$ as claimed.

In eq.~(\ref{eq:ds}), there exists a freedom of a global phase that cannot be determined by the field equation and does not seem to be important at first. However, when one constructs the field operators which consists of a linear combination of spinors, the choice of phases are no longer arbitrary since they are now relative. Up to an overall constant, the phase is determined by the demand of locality and Lorentz-invariance of the field operator. Additionally, assigning the spinors to the correct annihilation and creation operators are also non-trivial. Here, we will simply present the final solutions. For more details, please see~\cite[Sec.~5.5]{Weinberg:1995mt}. The Dirac field, with the appropriate normalization is given by
\begin{equation}
\psi(x)=(2\pi)^{-3}\int\frac{d^{3}p}{\sqrt{2E}}\sum_{\sigma}\left[e^{-ip\cdot x}u(\p,\sigma)a(\p,\sigma)
+e^{ip\cdot x}v(\p,\sigma)b^{\dag}(\p,\sigma)\right]
\end{equation}
where $a(\p,\sigma)$ and $b(\p,\sigma)$ satisfy the canonical anti-commutation relations
\begin{equation}
\{a(\p,\sigma),a^{\dag}(\p',\sigma')\}=\{b(\p,\sigma),b^{\dag}(\p',\sigma')\}=(2\pi)^{3}\delta(\p-\p')\delta_{\sigma\sigma'}.
\end{equation}
In the helicity basis, the spinors are given by\footnote{There is a subtle difference between the Dirac field in the helicity and polarization basis. In the latter, the spin does not have to be aligned to the momentum. As a result, when one integrates over the momentum to obtain the field operator, it includes helicity eigenstates as well as states whose spins are not aligned to the momentum. On the other hand, in the helicity basis, when one integrates over the momentum, only the helicity eigenstates are included.}
\begin{equation}
u(\p,\sigma)=\left[\begin{matrix}
\rho(\p,\sigma) \\
\rho(-\p,\sigma)
\end{matrix}\right],\quad
v(\p,\sigma)=(-1)^{1/2-\sigma}\left[\begin{matrix}
\rho(\p,-\sigma) \\
-\rho(-\p,-\sigma)
\end{matrix}\right].
\end{equation}

In the context of Elko to be constructed below, it is instructive to note that the left- and right-handed component of the Dirac spinors are constrained by the field equation. Taking the Dirac spinors to be
\begin{equation}
\psi(\p,\sigma)\equiv\left[\begin{matrix}
\psi_{R}(\p,\sigma) \\
\psi_{L}(\p,\sigma)
\end{matrix}\right]
\end{equation}
one finds that
\begin{equation}
\left[\begin{matrix}
\psi_{R}(\p,\sigma) \\
\psi_{L}(\p,\sigma)
\end{matrix}\right]=\pm
\left[\begin{matrix}
O & (\bar{\sigma}^{\mu}p_{\mu}/m) \\
(\sigma^{\mu}p_{\mu}/m) & O
\end{matrix}\right]
\left[\begin{matrix}
\psi_{R}(\p,\sigma) \\
\psi_{L}(\p,\sigma)
\end{matrix}\right]
\end{equation}
where $\sigma^{\mu}=(I,\s)$ and $\bar{\sigma}^{\mu}(I,-\s)$. Therefore, the Dirac spinors can also be written as
\begin{equation}
\psi(\p,\sigma)=\left[\begin{matrix}
\pm(\bar{\sigma}^{\mu}p_{\mu}/m)\psi_{L}(\p,\sigma) \\
\psi_{L}(\p,\sigma)
\end{matrix}\right]=
\left[\begin{matrix}
\psi_{R}(\p,\sigma) \\
\pm(\sigma^{\mu}p_{\mu}/m)\psi_{R}(\p,\sigma)
\end{matrix}\right]. \label{eq:psirl}
\end{equation}

\subsection{Elko}\label{ek}

We have shown in the previous section how the Dirac spinors can be obtained by relating the $(\frac{1}{2},0)$ and $(0,\frac{1}{2})$ representation via parity. By applying a different operation, Elko can also be obtained in a similar manner. There are two equivalent, but slightly differently approaches to construct Elko. We will start with the original one given by Ahluwalia and Grumiller~\cite{Ahluwalia:2004ab,Ahluwalia:2004sz}, utilizing the Wigner time-reversal matrix $\Theta$ which satisfies the identity
\begin{equation}
\Theta\s\Theta^{-1}=-\s^{*}
\end{equation}
from which we find that $\vartheta\Theta\chi^{*}(\e,\sigma)$ where $\vartheta$ is a phase to be determined, transforms as a right-handed spinor
\begin{equation}
\vartheta\Theta\chi^{*}(\p,\sigma)=\exp\left(\frac{1}{2}\s\cdot\bv\right)\left[\vartheta\Theta\chi^{*}(\e,\sigma)\right].
\end{equation}
Therefore, we can construct a four-component spinor of the form
\begin{equation}
\lambda(\p,\sigma)=\left[\begin{matrix}
\vartheta\Theta\chi^{*}(\p,\sigma) \\
\chi(\p,\sigma)
\end{matrix}\right].
\end{equation}
Elko is obtained by setting the phase to $\vartheta=\pm i$. They are eigenspinors of the charge-conjugation operator
\begin{equation}
\mathcal{C}=\left(\begin{matrix}
O & -i\Theta \\
i\Theta & O
\end{matrix}\right)K
\end{equation}
where $K$ is the complex conjugation operator that acts to its right. The operator $\mathcal{C}$ is identified with charge-conjugation by virtues of its action on the Dirac spinors
\begin{equation}
\mathcal{C}u(\p,\sigma)=iv(\p,\sigma),\quad
\mathcal{C}v(\p,\sigma)=iu(\p,\sigma).
\end{equation}
An explicit calculation yields
\begin{equation}
\mathcal{C}\lambda(\p,\sigma)\vert_{\vartheta=\pm i}=\pm\lambda(\p,\sigma)\vert_{\vartheta=\pm i}.
\end{equation}
Th spinors with positive and negative eigenvalues are called self-conjugate and anti-self-conjugate spinors respectively.

Elko has an important property called \textit{dual helicity} where the left- and right-handed component of $\lambda(\p,\sigma)$ have opposite helicity eigenvalues
\begin{equation}
\left(\frac{1}{2}\s\cdot\e\right)\left[\vartheta\Theta\chi^{*}(\e,\sigma)\right]=\mp\sigma
\left[\vartheta\Theta\chi^{*}(\e,\sigma)\right].
\end{equation}
In fact, it is related to $\chi(\e,-\sigma)$ by the identity
\begin{equation}
\Theta\chi^{*}(\e,\sigma)=(-1)^{1/2-\sigma}\chi(\e,-\sigma). \label{eq:opp}
\end{equation}
Equation~(\ref{eq:opp}) presents the second approach to construct Elko. Given a left-handed Weyl spinor $\chi(\e,\sigma)$ in the helicity basis, it follows that $\chi(\e,-\sigma)$ transforms as a right-handed Weyl spinor. In other words, the Dirac spinors utilizes parity $\p\rightarrow-\p$ while Elko utilizes helicity flip $\sigma\rightarrow-\sigma$. This construct allows us to draw comparison with its Dirac counterpart but it is only applicable in the helicity and not the polarization basis. The original construct based on the Wigner time-reversal operator does not have this limitation and is applicable in both basis.

A direct comparison with eq.~(\ref{eq:psirl}) is also possible. Firstly, we note that by virtue of eq.~(\ref{eq:hel}) and
\begin{equation}
\exp\left(-\frac{1}{2}\s\cdot\bv\right)=\sqrt{\frac{E+m}{2m}}\left[I-\frac{\s\cdot\p}{E+m}\right],
\end{equation}
we may write $\chi(\p,\sigma)$ as
\begin{equation}
\chi(\p,\sigma)=\sqrt{\frac{E+m}{2m}}\left[I-\frac{\sigma|\p|}{E+m}\right]\chi(\e,\sigma)
\end{equation}
so that
\begin{equation}
\Theta\chi^{*}(\p,\sigma)=\sqrt{\frac{E+m}{2m}}\left[I-\frac{\sigma|\p|}{E+m}\right](-1)^{1/2-\sigma}\chi(\e,-\sigma)\label{eq:tc}.
\end{equation}
Secondly, the helicity flip on $\chi(\e,\sigma)$ can also be achieved by introducing the following matrix
\begin{equation}
g(\phi)\equiv i\left(\begin{matrix}
0 & -e^{-i\phi} \\
e^{i\phi} & 0
\end{matrix}\right)
\end{equation}
where
\begin{equation}
g(\phi)\chi(\e,\sigma)=i(-1)^{1/2-\sigma}\chi(\e,-\sigma).\label{eq:gc}
\end{equation}
Using eqs.~(\ref{eq:tc}) and (\ref{eq:gc}), we obtain
\begin{equation}
g(\phi)\chi(\p,\sigma)=i\Theta\chi^{*}(\p,\sigma)
\end{equation}
which allows us to write Elko as
\begin{equation}
\lambda(\p,\sigma)=\left[\begin{matrix}
-i\vartheta g(\phi)\chi(\p,\sigma) \\
\chi(\p,\sigma)
\end{matrix}\right]. \label{eq:elko2}
\end{equation}
In this form, the difference between Dirac spinors and Elko can be understood from the different constraints imposed on the left- and right-handed spinors. From eq.~(\ref{eq:elko2}), we also obtain an identity for $\lambda(\p,\sigma)$
\begin{equation}
\mathcal{G}(\phi)\lambda(\p,\sigma)=\pm\lambda(\p,\sigma) \label{eq:gl}
\end{equation}
where
\begin{equation}
\mathcal{G}(\phi)\equiv\left[\begin{matrix}
O & g(\phi) \\
g(\phi) & O
\end{matrix}\right].
\end{equation}
This is analogous to the Dirac equation with $(\gamma^{\mu}p_{\mu}/m)$ replaced by $\mathcal{G}(\phi)$. But since $\mathcal{G}(\phi)$ has no energy dependence, eq.~(\ref{eq:gl}) contains no dynamics and cannot be regarded as a field equation for Elko. The introduction of $\mathcal{G}(\phi)$ may seem arbitrary, but it is in fact a defining feature of the theory. But as we will show in the latter sections, it appears in the spin-sums and propagator. The matrix, being non Lorentz-covariant, also characterizes the effects of Lorentz-violation.

From the apparent difference between $\psi(\p,\sigma)$ and $\lambda(\p,\sigma)$ and the fact that they are both constructed using the same Lorentz generators, it is simple to show that Elko does not satisfy the Dirac equation. For this purpose, we write down the solutions for the field operator to be constructed~\cite{Ahluwalia:2008xi,Ahluwalia:2009rh}
\begin{equation}
\lambda^{S}(\p,\sigma)=\left[\begin{matrix}
i\Theta\chi^{*}(\p,\sigma) \\
\chi(\p,\sigma)
\end{matrix}\right],\quad
\lambda^{A}(\p,\sigma)=(-1)^{1/2-\sigma}\left[\begin{matrix}
-i\Theta\chi^{*}(\p,-\sigma) \\
\chi(\p,-\sigma)
\end{matrix}\right]
\end{equation}
where the subscripts $S$ and $A$ stand for self-conjugate and anti-self-conjugate respectively. The phases and labels of $\lambda^{S,A}(\p,\sigma)$ are chosen to ensure the locality of the field operators. Acting the Dirac operator $\gamma^{\mu}p_{\mu}$ on Elko yields
\begin{eqnarray}
&& \gamma^{\mu}p_{\mu}\lambda^{S}(\p,\sigma)=im(-1)^{1/2-\sigma}\lambda^{S}(\p,-\sigma),\label{eq:gpl0}\\
&& \gamma^{\mu}p_{\mu}\lambda^{A}(\p,\sigma)=im(-1)^{-1/2+\sigma} \lambda^{A}(\p,-\sigma).
 \label{eq:gpl}
\end{eqnarray}
Applying $\gamma^{\mu}p_{\mu}$  from the left again on eqs.~(\ref{eq:gpl0}) and (\ref{eq:gpl}) yields the Klein-Gordon equation.  This  should not however be viewed as a derivation of the field equation since the Klein-Gordon equation is simply a statement of the energy-momentum relation. A proper analysis for the kinematics will be presented in sec.~\ref{mdof} where we perform canonical quantization.
\subsection{Elko dual and spin-sums}

We now come to the most important part of the theory - \textit{defining the Elko dual}. To see why this is necessary, we start with the usual Dirac dual namely\footnote{Whenever no possible confusion arises, we  use $\lambda(\p,\sigma)$ to represent both the self-conjugate and anti-self-conjugate Elko.}
\begin{equation}
\overline{\lambda}(\p,\sigma)=\lambda^{\dag}(\p,\sigma)\Gamma,\quad
\Gamma=\left(\begin{matrix}
O & I \\
I & O 
\end{matrix}\right).
\end{equation}
An explicit evaluation of the inner-product shows that the norm identically vanishes
\begin{equation}
\overline{\lambda}(\p,\sigma)\lambda(\p,\sigma)=0.
\end{equation}
To construct an orthonormal system for Elko, a new dual is needed.  For this purpose, we note that the Elko inner-products satisfy
\begin{eqnarray}
&&\overline{\lambda}^{S}(\p,\sigma)\lambda^{S}(\p,-\sigma')=+im(-1)^{1/2-\sigma}\delta_{\sigma\sigma'},\\
&&\overline{\lambda}^{A}(\p,\sigma)\lambda^{A}(\p,-\sigma')=-im(-1)^{1/2-\sigma}\delta_{\sigma\sigma'}.
\end{eqnarray}
Therefore, we define a new dual\footnote{The Elko dual initially introduced by Ahluwalia and Grumilelr reads $\dual{\lambda}(\p,\sigma)$~\cite{Ahluwalia:2004ab,Ahluwalia:2004sz}. Here, we adopt the convention in ref.~\cite{Ahluwalia:2016rwl} and will reserve the symbol $\neg$ for later purposes. Further details on the dual can be found in  refs.~\cite{Speranca:2013hqa,daSilvaa:2019kkt,Villalobos:2019fiu}.}
\begin{equation}
\widetilde{\lambda}(\p,\sigma)\equiv -i(-1)^{1/2-\sigma}\overline{\lambda}(\p,-\sigma)
\end{equation}
which forms an orthonormal
\begin{equation}
\widetilde{\lambda}^{S}(\p,\sigma)\lambda^{S}(\p,\sigma')=-\widetilde{\lambda}^{A}(\p,\sigma)\lambda^{A}(\p,\sigma')=m\delta_{\sigma\sigma'}
\end{equation}
and complete system
\begin{equation}
\frac{1}{m}\sum_{\sigma}\left[\lambda^{S}(\p,\sigma)\widetilde{\lambda}^{S}(\p,\sigma)-
\lambda^{A}(\p,\sigma)\widetilde{\lambda}^{A}(\p,\sigma)\right]=I
\end{equation}
just like their Dirac counterparts. However, their spin-sums are different. For definitiveness, the spin-sums under the Dirac and Elko dual are given by
\begin{eqnarray}
&& \sum_{\sigma}\lambda^{S}(\p,\sigma)\overline{\lambda}^{S}(\p,\sigma)=\frac{1}{2}\gamma^{\mu}p_{\mu}\left[I+\mathcal{G}(\phi)\right], \\
&& \sum_{\sigma}\lambda^{A}(\p,\sigma)\overline{\lambda}^{A}(\p,\sigma)=\frac{1}{2}\gamma^{\mu}p_{\mu}\left[I-\mathcal{G}(\phi)\right] 
\end{eqnarray}
and
\begin{eqnarray}
&& \sum_{\sigma}\lambda^{S}(\p,\sigma)\widetilde{\lambda}^{S}(\p,\sigma)=\frac{m}{2}\left[\mathcal{G}(\phi)+I\right],\label{eq:ss3} \\
&& \sum_{\sigma}\lambda^{A}(\p,\sigma)\widetilde{\lambda}^{S}(\p,\sigma)=\frac{m}{2}\left[\mathcal{G}(\phi)-I\right].\label{eq:ss4}
\end{eqnarray}

The Elko dual $\widetilde{\lambda}(\p,\sigma)$ gives us an orthonormal and complete system. But owing to the $\mathcal{G}(\phi)$ matrix, the theory is Lorentz-violating. 
The non-covariance can be removed using a new dual proposed by Ahluwalia~\cite{Ahluwalia:2016rwl}. The new dual is based on the observation that the matrix $\pm I+\tau\mathcal{G}(\phi)$, where $\tau$ is an arbitrary parameter has the following inverse~\cite{Lee:2014opa}
\begin{equation}
\left[\pm I+\tau\mathcal{G}(\phi)\right]^{-1}=\frac{\pm I-\tau\mathcal{G}(\phi)}{1-\tau^{2}}.
\end{equation}
Therefore, the Elko spin-sums given in eqs.~(\ref{eq:ss3}) and (\ref{eq:ss4}) are not invertible due to the poles at $\tau\pm1$. But by applying an infinitesimal deformation, replacing $\mathcal{G}(\phi)$ with $\tau\mathcal{G}(\phi)$, the spin-sums become invertible. Using the deformed spin-sums, we may introduce a new set of duals for Elko
\begin{equation}
\dual{\lambda}^{S}(\p,\sigma)\equiv\widetilde{\lambda}^{S}(\p,\sigma)\mathcal{A},\quad
\dual{\lambda}^{A}(\p,\sigma)\equiv\widetilde{\lambda}^{A}(\p,\sigma)\mathcal{B}
\end{equation}
where
\begin{eqnarray}
\mathcal{A}=\left[\frac{I-\tau\mathcal{G}(\phi)}{1-\tau^{2}}\right],\quad
\mathcal{B}=\left[\frac{I+\tau\mathcal{G}(\phi)}{1-\tau^{2}}\right]
\end{eqnarray}
such that the resulting spin-sums are Lorentz-invariant in the limit $\tau\rightarrow1$. For example, the spin-sum for the self-conjugate spinors are
\begin{eqnarray}
\sum_{\sigma}\lambda^{S}(\p,\sigma)\dual{\lambda}^{S}(\p,\sigma)&=&
\sum_{\sigma}\lambda^{S}(\p,\sigma)\left\{\widetilde{\lambda}^{S}(\p,\sigma)
\left[\frac{I-\tau\mathcal{G}(\phi)}{1-\tau^{2}}\right]\right\}\nonumber\\
&\equiv&\left[\sum_{\sigma}\lambda^{S}(\p,\sigma)\widetilde{\lambda}^{S}(\p,\sigma)\right]^{(\tau)}
\left[\frac{I-\tau\mathcal{G}(\phi)}{1-\tau^{2}}\right]\nonumber\\
&=&\frac{m}{2}\left[\frac{I+\tau\mathcal{G}(\phi)}{1-\tau^{2}}\right]\left[\frac{I-\tau\mathcal{G}(\phi)}{1-\tau^{2}}\right]\nonumber\\
&=&\left(\frac{m}{2}\right)I.\label{eq:dspinsum}
\end{eqnarray}
An important point is that in evaluating the spin-sums, the order of operation matters. While the use of $\tau$-deformation to define the inverses for $[\pm I+\tau\mathcal{G}(\phi)]$ can be justified via the theory of Moore and Penrose as shown in ref.~\cite{Bueno_Rogerio_2017}, the spin-sum and its deformation must be applied before multiplying it from the right by their pseudo-inverses as shown on the second line of eq.~(\ref{eq:dspinsum}).  That is, the order of operator is \textit{non-associative}. Whether this choice of ordering is mathematically justified needs further investigation. In light of the deformation, a subtle issue in the evaluation of Elko norm arises. To see this, we note that the inner-product between two spinors is equivalent to taking the trace of a term that contributes to the spin-sum
\begin{equation}
\dual{\lambda}(\p,\sigma)\lambda(\p,\sigma')=
\mbox{tr}\left[\lambda(\p,\sigma')\dual{\lambda}(\p,\sigma)\right].\label{eq:normtrace}
\end{equation}
Therefore, to evaluate eq.~(\ref{eq:normtrace}), we must deform  $\lambda(\p,\sigma')\dual{\lambda}(\p,\sigma)$. Using the identities
\begin{equation}
\lambda^{S}(\p,\sigma)=\left[\begin{matrix}
g\chi(\p,\sigma) \\
\chi(\p,\sigma)
\end{matrix}\right],\quad \quad
\lambda^{A}(\p,\sigma)=(-1)^{1/2-\sigma}\left[\begin{matrix}
-g\chi(\p,-\sigma) \\
\chi(\p,-\sigma)
\end{matrix}\right]
\end{equation}
the deformations are given by
\begin{eqnarray}
\lambda^{S}(\p,\sigma)\widetilde{\lambda}^{S}(\p,\sigma)\rightarrow\left[\lambda^{S}(\p,\sigma)\widetilde{\lambda}^{S}(\p,\sigma)\right]^{(\tau)}\equiv i(-1)^{1/2+\sigma}
\left[\begin{matrix}
g\chi(\p,\sigma)\chi^{\dag}(\p,-\sigma) & \tau g\chi(\p,\sigma)\chi^{\dag}(\p,-\sigma)g \\
\tau \chi(\p,\sigma)\chi^{\dag}(\p,-\sigma) & \chi(\p,\sigma)\chi^{\dag}(\p,-\sigma)g
\end{matrix}\right],\nonumber\\\label{eq:tt1}
\end{eqnarray}
\begin{eqnarray}
\lambda^{A}(\p,\sigma)\widetilde{\lambda}^{A}(\p,\sigma)\rightarrow\left[\lambda^{A}(\p,\sigma)\widetilde{\lambda}^{A}(\p,\sigma)\right]^{(\tau)}\equiv i
\left[\begin{matrix}
g\chi(\p,-\sigma)\chi^{\dag}(\p,\sigma) & -\tau g\chi(\p,-\sigma)\chi^{\dag}(\p,\sigma)g \\
-\tau \chi(\p,-\sigma)\chi^{\dag}(\p,\sigma) & \chi(\p,-\sigma)\chi^{\dag}(\p,\sigma)g
\end{matrix}\right].\nonumber\\\label{eq:tt2}
\end{eqnarray}
A straightforward calculation shows that the norms remain unchanged
\begin{equation}
\dual{\lambda}^{S}(\p,\sigma)\lambda^{S}(\p,\sigma')=-\dual{\lambda}^{A}(\p,\sigma)\lambda^{A}(\p,\sigma')=m\delta_{\sigma\sigma'}. \label{eq:nn}
\end{equation}

\section{Mass dimension one fermions} \label{mdof}

In the previous section, we have obtained all the necessary ingredients to construct a quantum field. The mass dimension one fermionic field, with the appropriate normalization is given by
\begin{equation}
\mathfrak{f}(x)=(2\pi)^{-3}\int\frac{d^{3}p}{\sqrt{2mE}}\sum_{\sigma}
\left[e^{-ip\cdot x}\lambda^{S}(\p,\sigma)a(\p,\sigma)+e^{ip\cdot x}\lambda^{A}(\p,\sigma)b^{\ddag}(\p,\sigma)\right].\label{eq:f}
\end{equation}
For its dual, there are two possibilities based on $\widetilde{\lambda}(\p,\sigma)$ and $\dual{\lambda}(\p,\sigma)$. Here, we will focus on the dual obtained from $\dual{\lambda}(\p,\sigma)$ and present the results for $\widetilde{\lambda}(\p,\sigma)$ in app.~\ref{ndf}. The dual of $\mathfrak{f}(x)$ is given by
\begin{equation}
\gdual{\mathfrak{f}}(x)=(2\pi)^{-3}\int\frac{d^{3}p}{\sqrt{2mE}}\sum_{\sigma}\left[e^{-ip\cdot x}\dual{\lambda}^{S}(\p,\sigma)a^{\ddag}(\p,\sigma)+e^{ip\cdot x}\dual{\lambda}^{A}(\p,\sigma)b(\p,\sigma)\right].\label{eq:df}
\end{equation}
The annihilation and creation operators for particles and anti-particles satisfy the canonical anti-commutation relations
\begin{equation}
\{a(\p,\sigma),a^{\ddag}(\p',\sigma')\}=\{b(\p,\sigma),b^{\ddag}(\p',\sigma')\}=(2\pi)^{3}\delta^{3}(\p-\p')\delta_{\sigma\sigma'}
\end{equation}
where all other anti-commutators identically vanish.

In eqs.~(\ref{eq:f}) and (\ref{eq:df}), we have used the symbol $\ddag$ instead of Hermitian conjugation for the creation operators. There is one main reason for this. Presently, we have an incomplete knowledge of the underlying symmetry for the theory. The properties of Elko have given us a few clues. One of the most important  being the non-trivial dual in the form of $\widetilde{\lambda}(\p,\sigma)$ and $\dual{\lambda}(\p,\sigma)$, neither of which can be directly obtained by Hermitian conjugation. This observation tentatively suggests that the adjoint for the operators and states may also be different though we do not know all of its properties. Nevertheless, for $\ddag$ to have the usual physical interpretation, it should satisfy
\begin{equation}
a^{\ddag\ddag}(\p,\sigma)\equiv a(\p,\sigma),\quad
\left[a^{\ddag}(\p,\sigma)b(\p',\sigma')\right]^{\ddag}\equiv b^{\ddag}(\p',\sigma')a(\p,\sigma).\label{eq:ddab}
\end{equation}

Having identified the appropriate field operator and dual associated with Elko, it is straightforward to obtain the Lagrangian and carry out canonical quantization. Towards this end, we first note that $\gdual{\mathfrak{f}}(x)$ satisfies an important property that its counterpart $\overline{\mathfrak{f}}(x)$ does not,  it anti-commutes with $\mathfrak{f}(x)$ at space-like separation
\begin{equation}
\{\mathfrak{f}(t,\x),\gdual{\mathfrak{f}}(t,\y)\}=O.
\end{equation}
This ensures that the Lagrangian as well as interactions constructed from $\mathfrak{f}(x)$ and $\gdual{\mathfrak{f}}(x)$ are local and preserves causality.

The Lagrangian can be directly inferred from the propagator which is obtained by simply computing the two-point time-ordered product $\langle\,\,|\mathcal{T}[\mathfrak{f}(x)\gdual{\mathfrak{f}}(y)]|\,\,\rangle$. Using the deformed Lorentz-invariant spin-sums, we obtain a Klein-Gordon propagator
\begin{equation}
S(x-y)=\frac{i}{2}\int \frac{d^{4}q}{(2\pi)^{4}}\,e^{-iq\cdot(x-y)}\frac{I}{q^{2}-m^{2}+i\epsilon}
\end{equation}
which shows that the field and its dual are indeed of mass dimension one. Therefore, the Lagrangian is given by
\begin{equation}
\mathscr{L}=\partial^{\mu}\gdual{\mathfrak{f}}\partial_{\mu}\mathfrak{f}-m^{2}\gdual{\mathfrak{f}}\mathfrak{f}.
\end{equation}
The canonical momentum is simply $\mathfrak{p}(x)=\partial\gdual{\mathfrak{f}}(x)/\partial t$
and they satisfy the canonical anti-commutation relations
\begin{eqnarray}
&& \{\mathfrak{f}(t,\x),\mathfrak{f}(t,\y)\}=\{\pi(t,\x),\pi(t,\y)\}=O,\\
&& \{\mathfrak{f}(t,\x),\pi(t,\y)\}=\left(\frac{i}{2}\right)\delta^{3}(\x-\y)I.
\end{eqnarray}

The Lagrangian has another important physical property - it yields the correct free energy-momentum tensor. Under an infinitesimal space-time translation $\epsilon^{\mu}(x)$, we have
\begin{equation}
\delta\mathfrak{f}=\epsilon^{\mu}\partial_{\mu}\mathfrak{f},\quad
\delta\gdual{\mathfrak{f}}=\epsilon^{\mu}\partial_{\mu}\gdual{\mathfrak{f}}. \label{eq:de}
\end{equation}
Therefore, the variation on the action is
\begin{eqnarray}
\delta S&=&\int d^{4}x\left[
\delta\gdual{\mathfrak{f}}\frac{\delta\mathscr{L}}{\partial \df}
+\delta(\partial_{\mu}\gdual{\mathfrak{f}})\frac{\partial\mathscr{L}}{\partial(\partial_{\mu}\df)}
+\frac{\delta\mathscr{L}}{\partial \mathfrak{f}}\delta\mathfrak{f}
+\frac{\partial\mathscr{L}}{\partial(\partial_{\mu}\mathfrak{f})}\label{eq:dI}
\delta(\partial_{\mu}\mathfrak{f})\right]\nonumber\\
&=&\int d^{4}x\left[\epsilon^{\mu}\partial_{\mu}\mathscr{L}
+(\partial_{\mu}\epsilon^{\nu})(\partial_{\nu}\gdual{\mathfrak{f}})\frac{\partial\mathscr{L}}{\partial(\partial_{\mu}\df)}
+\frac{\partial\mathscr{L}}{\partial(\partial_{\mu}\mathfrak{f})}(\partial_{\mu}\epsilon^{\nu})
(\partial_{\nu}\mathfrak{f})
\right].
\end{eqnarray}
Integrating by parts and demanding $\delta S=0$, we obtain the energy-momentum tensor
\begin{eqnarray}
T^{\mu\nu}&=&-\eta^{\mu\nu}\mathscr{L}+
(\partial^{\nu}\gdual{\mathfrak{f}})\frac{\partial\mathscr{L}}{\partial(\partial_{\mu}\df)}
+\frac{\partial\mathscr{L}}{\partial(\partial_{\mu}\mathfrak{f})}
(\partial^{\nu}\mathfrak{f})\nonumber\\
&=&-\eta^{\mu\nu}\mathscr{L}+\partial^{\nu}\gdual{\mathfrak{f}}\partial^{\mu}\mathfrak{f}
+\partial^{\mu}\gdual{\mathfrak{f}}\partial^{\nu}\mathfrak{f}
\end{eqnarray}
so that the free Hamiltonian
\begin{equation}
H_{0}=\int d^{3}x\left(\partial_{t}\gdual{\mathfrak{f}}\partial_{t}\mathfrak{f}-\partial_{i}\gdual{\mathfrak{f}}\partial^{i}\mathfrak{f}+m^{2}\gdual{\mathfrak{f}}\mathfrak{f}\right).
\label{eq:H0}
\end{equation}
Substituting eqs.~(\ref{eq:f}) and (\ref{eq:df}) into $H_{0}$, we obtain
\begin{eqnarray}
H_{0}&=&\int d^{3}p\left(\frac{E}{m}\right)\sum_{\sigma\sigma'}\left[\dual{\lambda}^{S}(\p,\sigma)\lambda^{S}(\p,\sigma')a^{\ddag}(\p,\sigma)a(\p,\sigma')+\dual{\lambda}^{A}(\p,\sigma)\lambda^{A}(\p,\sigma')b(\p,\sigma)b^{\ddag}(\p,\sigma')\right].\nonumber\\
&=&\int d^{3}p\, E\sum_{\sigma}\left[a^{\ddag}(\p,\sigma)a(\p,\sigma)+b(\p,\sigma)b^{\ddag}(\p,\sigma)\right].
\label{eq:h0}
\end{eqnarray}
Similarly, it is straightforward to show that
\begin{eqnarray}
P^{i}_{0}=\int d^{3}p\, p^{i}\sum_{\sigma}\left[a^{\ddag}(\p,\sigma)a(\p,\sigma)+b(\p,\sigma)b^{\ddag}(\p,\sigma)\right].
\end{eqnarray}

Our evaluation of $P^{\mu}_{0}$ reveals an interesting fact. In general, when the expansion coefficients for the field operator and its dual form an orthonormal system, a Klein-Gordon-like Lagrangian will always yield a physical energy-momentum operator. Of course, this result is only a necessary but not sufficient condition. The Lagrangian must also yield a set of local commutators/anti-commutators at space-like separation. For instance, a Klein-Gordon Lagrangian for the Dirac field would indeed give us a physical $P^{\mu}_{0}$, but the resulting field-momentum anti-commutator is non-local. Additionally, a Klein-Gordon Lagrangian would also be inconsistent with the propagator obtained from evaluating $\langle\,\,|\mathcal{T}\psi(x)\overline{\psi}(y)|\,\,\rangle$. In a nutshell, while the Klein-Gordon Lagrangian for $\mathfrak{f}(x)$ and $\gdual{\mathfrak{f}}(x)$ appears to be trivial, there is an intricate underlying structure that ensures the self-consistency of the theory.

Upon introducing $\df(x)$, there is another important issue that we have to address. The Lagrangian we have constructed is not real. A related issue is that if $\ddag$ is not Hermitian conjugation, then the Hamiltonian is not Hermitian. Nevertheless, from eq.~(\ref{eq:h0}), the Hamiltonian still has a real physical spectrum and it also has the property $H^{\ddag}_{0}=H_{0}$. Therefore, instead of demanding the Lagrangian to be real, the property of the Hamiltonian tentatively suggests the condition $\mathscr{L}^{\ddag}=\mathscr{L}$. This condition can be satisfied by introducing the following operation on Elko
\begin{equation}
\lambda^{\ddag}(\p,\sigma)\equiv  -i(-1)^{1/2-\sigma}\lambda^{\dag}(\p,-\sigma) \label{eq:ddag}
\end{equation}
from which we may write $\widetilde{\lambda}(\p,\sigma)=\lambda^{\ddag}(\p,\sigma)\Gamma$ in an analogous form to the Dirac dual. Using eq.~(\ref{eq:ddag}), it is straightforward to show that 
\begin{eqnarray}
&& \lambda^{\ddag\ddag}(\p,\sigma)=\lambda(\p,\sigma), \label{eq:ddag1}\\
&& [\dual{\lambda}(\p,\sigma)\lambda(\p',\sigma')]^{\ddag}=\dual{\lambda}(\p',\sigma')\lambda(\p,\sigma) \label{eq:ddag2}
\end{eqnarray}
 and hence $\mathscr{L}^{\ddag}=\mathscr{L}$. It is tempting to extend the definition of eq.~(\ref{eq:ddag}) to $a^{\ddag}(\p,\sigma)$ and $b^{\ddag}(\p,\sigma)$. However, such an extension maybe going a step too far and for now, it is beyond the scope of this paper. Apart from the basic properties of $\ddag$ given in eq.~(\ref{eq:ddab}) that ensures a physical particle interpretation, it would be desirable to determine other properties from the underlying symmetries of the theory. It has been noted in ref.~\cite{Speranca:2013hqa} that eq.~(\ref{eq:ddag}) can also be written as
\begin{equation}
\lambda^{\ddag}(\p,\sigma)=\left[\Xi(\p)\lambda(\p,\sigma)\right]^{\dag}\label{eq:ddag3}
\end{equation}
where $\Xi(\p)=\mathcal{G}(\phi)\gamma^{\mu}p_{\mu}/m$. This is an elegant relation between $\lambda^{\ddag}(\p,\sigma)$ and $\lambda^{\dag}(\p,\sigma)$ that seems to render the former definition redundant. But since $\Xi(\p)$ is momentum-dependent, we find that by substituting eq.~(\ref{eq:ddag3}) into (\ref{eq:ddag2}), the equality is violated. In other words, the scope of applicability of $\Xi(\p)$ is limited. In our opinion, eq.~(\ref{eq:ddag3}) is an important algebraic identity that should be satisfied for all momentum. Therefore, one should view $\ddag$  as a general operation that is a composition of Hermitian conjugation and $\sigma\rightarrow-\sigma$.

The dual that we have introduced is not only important for the free theory, they also play an important role in our attempt to formulate a consistent interacting theory. For instance, demanding the Lagrangian to satisfy $\mathscr{L}=\mathscr{L}^{\ddag}$ would constrain the allowed interactions. There is also the issue on whether we should compute observables based on the standard $S$-matrix. These results will be presented elsewhere in a future publication.

\section{Higher-spin generalizations}\label{hs}

The theory of spin-half Elko and mass dimension one fermion have a natural generalization to the $(j,0)\oplus(0,j)$ representation~\cite{Lee:2012td}.\footnote{In the process of preparing this article, we have found mistakes in ref.~\cite{Lee:2012td} on the construction of bosonic fields. The corrected results are presented here.} The approach parallels the spin-half construct. This time, we start with a function $\chi_{j}(\e,\sigma)$ belonging to the $(j,0)$ representation in the helicity basis so that\footnote{We reserve the word 'spinor' strictly for the spin-half representation.}
\begin{equation}
(\J\cdot\e) \chi_{j}(\e,\sigma)=\sigma\chi_{j}(\e,\sigma)\label{eq:je}
\end{equation}
where $\J$ is the rotation generator of dimension $2j+1$ and $\sigma=-j,\cdots,j$. Their solutions are given by
\begin{eqnarray}
&& (J_{x}\pm iJ_{y})_{\sigma\bar{\sigma}}
=\delta_{\sigma,\bar{\sigma}\pm1}\sqrt{(j\mp\bar{\sigma})(j\pm\bar{\sigma}+1)}\label{eq:r12},\\
&& (J_{z})_{\sigma\bar{\sigma}}=\sigma\delta_{\sigma\bar{\sigma}}.\label{eq:r3}
\end{eqnarray}
Using the identity
\begin{equation}
\J^{*}_{\bar{\sigma}\sigma}=-(-1)^{\bar{\sigma}-\sigma}\J_{-\bar{\sigma}-\sigma},
\end{equation}
up to a sign, we obtain the Wigner time-reversal operator for spin-$j$
\begin{equation}
(\Theta_{j})_{\bar{\sigma}\sigma}=(-1)^{-j-\bar{\sigma}}\delta_{-\bar{\sigma}\sigma}
\end{equation}
where it satisfies
\begin{equation}
\Theta_{j}\J\Theta^{-1}_{j}=-\J^{*}.\label{eq:wtr}
\end{equation}
Here, the sign is fixed by demanding that when $j=\frac{1}{2}$, we obtain $\Theta=-i\sigma_{y}$. Therefore, we obtain a function of the form
\begin{equation}
\lambda_{j}(\p,\sigma)=\left[\begin{matrix}
\vartheta\Theta_{j}\chi^{*}_{j}(\p,\sigma) \\
\chi_{j}(\p,\sigma)
\end{matrix}\right].
\end{equation}
They become eigenfunctions of the charge-conjugation operator $\mathcal{C}_{j}$
\begin{equation}
\mathcal{C}_{j}=\left(\begin{matrix}
O & -i\Theta^{-1}_{j} \\
-i\Theta_{j} & O
\end{matrix}\right)K
\end{equation}
when the phase is fixed to $\vartheta=\pm i$. A straightforward calculation reveals
\begin{equation}
\mathcal{C}_{j}\lambda_{j}(\p,\sigma)\vert_{\vartheta=\pm i}=\mp (-1)^{2j}\lambda_{j}(\p,\sigma)\vert_{\vartheta=\pm i}.
\end{equation}

The explicit solutions of $\chi_{j}(\e,\sigma)$ can be directly obtained from eq.~(\ref{eq:je}) but the calculation becomes tedious for higher-spin. Here, it is more convenient to use the following relation~\footnote{While it is possible to construct local mass dimension one fields in the polarization basis, non-trivial phases containing information about the direction of boosts must be encoded in the rest spinors/functions~\cite{Ahluwalia:2009rh,Lee:2012td}. In our opinion, this is not entirely satisfactory so further investigation is required.}
\begin{equation}
\chi_{j}(\e,\sigma)=S\chi_{j}(\0,\sigma)\label{eq:h2p}
\end{equation}
where $\chi_{j}(\0,\sigma)$ is an eigenfunction of $J_{z}$ and $SJ_{z}S^{-1}=(\J\cdot\e)$. The matrix $S$ can be obtained by diagonalizing $\J\cdot\e$. However, by imposing some reasonable ansatz, we do not need the explicit solution of $S$ to evaluate the norms and spin-sums.

We take the self-conjugate and anti-self-conjugate functions to be
\begin{eqnarray}
\lambda^{S}_{j}(\0,\sigma)=\lambda_{j}(\0,\sigma)\vert_{\vartheta=i,\chi_{j}(\mathbf{0},\sigma)\rightarrow\omega(\mathbf{0},\sigma)},\quad
\lambda^{A}_{j}(\0,\sigma)=\lambda_{j}(\0,-\sigma)\vert_{\vartheta=-i,\chi_{j}(\mathbf{0},\sigma)\rightarrow\varepsilon(\mathbf{0},\sigma)}\nonumber\\
\end{eqnarray}
with
\begin{equation}
\omega_{\ell}(\0,\sigma)=c_{j}\sqrt{\frac{m}{2}}\delta_{\ell\sigma},\quad
\varepsilon_{\ell}(\0,\sigma)=d_{j}\sqrt{\frac{m}{2}}\delta_{\ell\sigma}
\end{equation}
where $c_{j}$ and $d_{j}$ are constants to be determined. The norms of the functions are basis- and momentum-independent.\footnote{Here, the functions $\chi_{j}(\p,\sigma)$ and $\lambda_{j}(\p,\sigma)$ are always in the helicity basis.} Like the spin-half construct, we find
\begin{equation}
\overline{\lambda}_{j}(\p,\sigma)\lambda_{j}(\p,\sigma)=0
\end{equation}
and the general inner-products are given by
\begin{eqnarray}
&& \overline{\lambda}^{S}_{j}(\p,\sigma)\lambda^{S}_{j}(\p,\sigma')=+\frac{im(-1)^{-j-\sigma}}{2}\left[c^{*2}_{j}-(-1)^{-2\sigma}c^{2}_{j}\right]\delta_{\sigma,-\sigma'},\label{eq:xi_dual}\\
&& \overline{\lambda}^{A}_{j}(\p,\sigma)\lambda^{A}_{j}(\p,\sigma')=-\frac{im(-1)^{-j-\sigma}}{2}
\left[d^{*2}_{j}-(-1)^{-2\sigma}d^{2}_{j}\right]\delta_{\sigma,-\sigma'}. \label{eq:zeta_dual}
\end{eqnarray}
Therefore, we define the dual
\begin{eqnarray}
&& \widetilde{\lambda}^{S}_{j}(\p,\sigma)\equiv\beta(\sigma)\lambda^{S\dag}_{j}(\p,-\sigma)\Gamma, \\
&& \widetilde{\lambda}^{A}_{j}(\p,\sigma)\equiv\beta(\sigma)\lambda^{A\dag}_{j}(\p,-\sigma)\Gamma
\end{eqnarray}
where $\beta(\sigma)$ is a phase to be determined. These inner-products satisfy the relation
\begin{equation}
\overline{\lambda}_{j}(\p,\sigma)\lambda_{j}(\p,-\sigma)=-\overline{\lambda}_{j}(\p,\sigma\pm 1)\lambda_{j}(\p,-\sigma\mp 1). \label{eq:Elko_dual}
\end{equation}
We would like the self-conjugate and anti-self-conjugate functions to have the same norm in their respective sectors so the phases are fixed to
\begin{equation}
\beta(\sigma)=-\beta(\sigma\pm1) \label{eq:recursive_beta}
\end{equation}
which can be rewritten as
\begin{equation}
\beta(\sigma)=(-1)^{j-\sigma}\beta(j). \label{eq:recursive_beta2}
\end{equation}
The norms now take the form
\begin{eqnarray}
&&\widetilde{\lambda}^{S}_{j}(\p,\sigma)\lambda^{S}_{j}(\p,\sigma')=
\begin{cases}
im\beta(j)(c^{*2}_{j}+c^{2}_{j})\delta_{\sigma\sigma'}/2 & j=\frac{1}{2},\frac{3}{2},\cdots \\
im\beta(j)(c^{*2}_{j}-c^{2}_{j})\delta_{\sigma\sigma'}/2 & j=1,2,\cdots \label{eq:xi_norms}
\end{cases}\\
&&\widetilde{\lambda}^{A}_{j}(\p,\sigma)\lambda^{A}_{j}(\p,\sigma')=
\begin{cases}
-im\beta(j)(d^{*2}_{j}+d^{2}_{j})\delta_{\sigma\sigma'}/2 & j=\frac{1}{2},\frac{3}{2},\cdots \\
-im\beta(j)(d^{*2}_{j}-d^{2}_{j})\delta_{\sigma\sigma'}/2 & j=1,2,\cdots. \label{eq:zeta_norms}
\end{cases}
\end{eqnarray}
Demanding the norms to be real, the constants squared must be either real or imaginary. We fix them to
\begin{equation}
c^{*2}_{j}=-(-1)^{2j}c^{2}_{j},\quad
d^{*2}_{j}=-(-1)^{2j}d^{2}_{j}\label{eq:cd_constants}
\end{equation}
to obtain
\begin{eqnarray}
&&\widetilde{\lambda}^{S}_{j}(\p,\sigma)\lambda^{S}_{j}(\p,\sigma')=im\beta(j)c^{*2}_{j}\delta_{\sigma\sigma'},\label{eq:nfs}\\
&&\widetilde{\lambda}^{A}_{j}(\p,\sigma)\lambda^{A}_{j}(\p,\sigma')=-im\beta(j)d^{*2}_{j}\delta_{\sigma\sigma'}.\label{eq:nfa}
\end{eqnarray}

Computing the spin-sums in the helicity basis requires a bit more work. We will first compute them in the polarization basis at rest before transforming to the helicity basis. We find
\begin{eqnarray}
&&\sum_{\sigma}\lambda^{S}_{j}(\0,\sigma)\widetilde{\lambda}^{S}_{j}(\0,\sigma)
=\frac{1}{2}\left[im\beta(j)c^{*2}_{j}\left(\begin{matrix}
I & O \\
O & I \end{matrix}\right)
+m\beta(j)|c_{j}|^{2}(-1)^{2j}\left(\begin{matrix}
O & \Theta_{j}\\
\Theta_{j} & O \end{matrix}\right)\right],\label{eq:N0}\\
&&\sum_{\sigma}\lambda^{A}_{j}(\0,\sigma)\widetilde{\lambda}^{A}_{j}(\0,\sigma)
=\frac{1}{2}\left[-im\beta(j)d^{*2}_{j}\left(\begin{matrix}
I & O \\
O & I \end{matrix}\right)
+m\beta(j)|d_{j}|^{2}(-1)^{2j}\left(\begin{matrix}
O & \Theta_{j}\\
\Theta_{j} & O \end{matrix}\right)\right].\nonumber\\ \label{eq:M0}
\end{eqnarray}
The spin-sums in the helicity basis are obtained by performing a similarity transformation using the direct sum of $S$. The results are
\begin{eqnarray}
\sum_{\sigma}\lambda^{S}_{j}(\e,\sigma)\widetilde{\lambda}^{S}_{j}(\e,\sigma)
&=&\frac{1}{2}\left[im\beta(j)c^{*2}_{j}\left(\begin{matrix}
I & O \\
O & I \end{matrix}\right)
+m\beta(j)|c_{j}|^{2}(-1)^{2j}
\left(\begin{matrix}
O & S\Theta_{j} S^{\dag} \\
S\Theta_{j} S^{\dag} & O \end{matrix}\right)\right],\nonumber\\
\end{eqnarray}
\begin{eqnarray}
\sum_{\sigma}\lambda^{S}_{j}(\e,\sigma)\widetilde{\lambda}^{S}_{j}(\e,\sigma)
&=&\frac{1}{2}\left[-im\beta(j)d^{*2}_{j}\left(\begin{matrix}
I & O \\
O & I \end{matrix}\right)
+m\beta(j)|d_{j}|^{2}(-1)^{2j}
\left(\begin{matrix}
O & S\Theta_{j} S^{\dag} \\
S\Theta_{j} S^{\dag} & O \end{matrix}\right)\right].\nonumber\\
\end{eqnarray}
The spin-sum at arbitrary momentum is given by
\begin{equation}
\sum_{\sigma}\lambda_{j}(\p,\sigma)\widetilde{\lambda}_{j}(\p,\sigma)=\exp(i\boldsymbol{\mathcal{K}}\cdot\bv)\left[\sum_{\sigma}\lambda_{j}(\e,\sigma)\widetilde{\lambda}_{j}(\e,\sigma)\right]\exp(-i\boldsymbol{\mathcal{K}}\cdot\bv)
\end{equation}
where the boost generator is given by
\begin{equation}
\boldsymbol{\mathcal{K}}=\left(\begin{matrix}
+i\J & O \\
O & -i\J
\end{matrix}\right).
\end{equation}
Using eq.~(\ref{eq:wtr}) and the relation $\exp(\J\cdot\bv)=S\exp(J_{z}\varphi)S^{-1}$, we find that the spin-sums in the helicity basis to be boost-independent
\begin{eqnarray}
&&\sum_{\sigma}\lambda^{S}_{j}(\p,\sigma)\widetilde{\lambda}^{S}_{j}(\p,\sigma)
=\frac{m}{2}\left[+i\beta(j)c^{*2}_{j}I
+|c|^{2}_{j}\mathcal{G}_{j}\right],\\
&&\sum_{\sigma}\lambda^{A}_{j}(\p,\sigma)\widetilde{\lambda}^{A}_{j}(\p,\sigma)
=\frac{m}{2}\left[-i\beta(j)d^{*2}_{j}I
+|d|^{2}_{j}\mathcal{G}_{j}\right]
\end{eqnarray}
where 
\begin{equation}
\mathcal{G}_{j}\equiv\left(\begin{matrix}
O & g_{j} \\
g_{j} & O
\end{matrix}\right),\quad g_{j}\equiv \beta(j)(-1)^{2j} S\Theta_{j} S^{-1}.
\end{equation}

The remaining task is to determine $g_{j}$ and the proportionality constants.  Towards this end, we shall use the $j=\frac{1}{2}$ spin-sums given in eqs.~(\ref{eq:ss3}) and (\ref{eq:ss4}) as the initial ansatz. Using eq.~(\ref{eq:wtr}) and $SJ_{z}S^{-1}=(\J\cdot\e)$, we obtain the condition
\begin{equation}
\{g_{j},\J\cdot\e\}=O. \label{eq:gja}
\end{equation}
From the form of $\mathcal{G}_{1/2}(\phi)$, we assume that $g_{j}$ is off-diagonal and is of the form
\begin{equation}
(g_{j})_{\ell m}=\beta(j)(-1)^{2j}f_{\ell}(\phi)\delta_{\ell,-m}.\label{eq:gj}
\end{equation}
Substituting the solutions of $\J$ and eq.~(\ref{eq:gj}) into (\ref{eq:gja}), we obtain
\begin{equation}
f_{\sigma}(\phi)=-f_{-\sigma+1}(\phi)e^{2i\phi}.
\end{equation}
When $j=\frac{1}{2}$, without the loss of generality, we make the following choices
\begin{equation}
f_{-1/2}(\phi)=e^{i\phi},\quad \beta(\textstyle{\frac{1}{2}})=-i,\quad c_{1/2}=d_{1/2}=1. \label{eq:bhalf}
\end{equation}
It is tempting to extrapolate the solutions for $\beta(\frac{1}{2})$, $c_{1/2}$ and $d_{1/2}$ to all spin which then according to eqs.~(\ref{eq:nfs}) and (\ref{eq:nfa}), would determine the norms of the functions. But such an extrapolation turns out to be incorrect. Instead, the correct choices that ensures locality, spin-statistics and positivity of the free Hamiltonian are
\begin{equation}
\beta_{j}=-i,\quad c_{j}=1,\quad d_{j}=(-1)^{2j}.
\end{equation}
Extrapolating the solution for $f_{-1/2}(\phi)$, we obtain
\begin{equation}
f_{\ell}(\phi)=(-1)^{j+\ell}e^{-2i\ell\phi},\quad \ell=-j,\cdots,j.
\end{equation}
Therefore, the matrix $g_{j}(\phi)$ is determined to be
\begin{equation}
(g_{j})_{\ell m}(\phi)=i(-1)^{j-\ell+1}e^{-2i\ell\phi}\delta_{\ell,-m}.
\end{equation}
For example, when $j=\frac{1}{2}$ and $j=1$, we have
\begin{equation}
g_{1/2}(\phi)=i\left(\begin{matrix}
0 & e^{-i\phi} \\
e^{i\phi} & 0
\end{matrix}\right),\quad
g_{1}(\phi)=-i\left(\begin{matrix}
0 & 0 & e^{-2i\phi} \\
0 & -1 & 0 \\
e^{2i\phi} & 0 & 0
\end{matrix}\right).
\end{equation}
The final form of the spin-sums are given by
\begin{eqnarray}
&&\sum_{\sigma}\lambda^{S}_{j}(\p,\sigma)\widetilde{\lambda}^{S}_{j}(\p,\sigma)
=\frac{m}{2}\left[I
+\mathcal{G}_{j}(\phi)\right],\label{eq:hs1}\\
&&\sum_{\sigma}\lambda^{A}_{j}(\p,\sigma)\widetilde{\lambda}^{A}_{j}(\p,\sigma)
=\frac{m}{2}\left[(-1)^{2j}I+\mathcal{G}_{j}(\phi)\right].\label{eq:hs2}
\end{eqnarray}
Similar to the spin-half construct, using the generic property $\mathcal{G}^{2}_{j}(\phi)=I$, we may introduce a new set of duals
\begin{equation}
\dual{\lambda}^{S}_{j}(\p,\sigma)\equiv\widetilde{\lambda}^{S}_{j}(\p,\sigma)\mathcal{A}_{j},\quad
\dual{\lambda}^{A}_{j}(\p,\sigma)\equiv\widetilde{\lambda}^{A}_{j}(\p,\sigma)\mathcal{B}_{j}
\end{equation}
with
\begin{equation}
\mathcal{A}_{j}=\frac{I-\tau\mathcal{G}_{j}(\p)}{1-\tau^{2}},\quad
\mathcal{B}_{j}=\frac{I-(-1)^{2j}\tau\mathcal{G}_{j}(\p)}{1-\tau^{2}}
\end{equation}
so that their spin-sums are
\begin{eqnarray}
&& \sum_{\sigma}\lambda^{S}_{j}(\p,\sigma)\dual{\lambda}^{S}_{j}(\p,\sigma)=\left(\frac{m}{2}\right)I,\\
&& \sum_{\sigma}\lambda^{A}_{j}(\p,\sigma)\dual{\lambda}^{A}_{j}(\p,\sigma)=\left(\frac{m}{2}\right)(-1)^{2j}I.
\end{eqnarray}

\subsection{Locality structures and Hamiltonians} \label{lsh}
Having determined the spin-sums and norms, it is straightforward to construct the field operators and their duals. Using $\lambda_{j}(\p,\sigma)$ and $\dual{\lambda}_{j}(\p,\sigma)$, we obtain
\begin{eqnarray}
&&\mathfrak{f}_{j}(x)=(2\pi)^{-3}\int\frac{d^{3}p}{\sqrt{2mE}}\sum_{\sigma}
\left[e^{-ip\cdot x}\lambda^{S}(\p,\sigma)a(\p,\sigma)+e^{ip\cdot x}\lambda^{A}(\p,\sigma)b^{\ddag}(\p,\sigma)\right],\\
&&\df_{j}(x)=(2\pi)^{-3}\int\frac{d^{3}p}{\sqrt{2mE}}\sum_{\sigma}
\left[e^{ip\cdot x}\dual{\lambda}^{S}(\p,\sigma)a^{\ddag}(\p,\sigma)+e^{-ip\cdot x}\dual{\lambda}^{A}(\p,\sigma)b(\p,\sigma)\right].
\end{eqnarray}
By virtues of the spin-sums, we find
\begin{equation}
[\mathfrak{f}_{j}(t,\x),\df_{j}(t,\y)]_{\pm}=O.
\end{equation}
The propagator and Lagrangian are given by
\begin{eqnarray}
&& S_{j}(x-y)=\frac{i}{2}\int \frac{d^{4}q}{(2\pi)^{4}}\,e^{-iq\cdot(x-y)}\frac{I}{q^{2}-m^{2}+i\epsilon},\\
&& \mathscr{L}_{j}=\partial^{\mu}\df_{j}\partial_{\mu}\mathfrak{f}_{j}-m^{2}\df_{j}\mathfrak{f}_{j}.
\end{eqnarray}
The canonical quantization is straightforward. Taking $\mathfrak{p}_{j}(x)=\partial\df_{j}(x)/\partial t$, we obtain the standard results with the usual spin-statistics for bosons and fermions
\begin{eqnarray}
&& [\mathfrak{f}_{j}(t,\x),\mathfrak{f}_{j}(t,\y)]_{\pm}=[\mathfrak{p}_{j}(t,\x),\mathfrak{p}_{j}(t,\y)]_{\pm}=O,\\
&& [\mathfrak{f}_{j}(t,\x),\mathfrak{p}_{j}(t,\y)]_{\pm}=\left(\frac{i}{2}\right)\delta^{3}(\x-\y)I.
\end{eqnarray}
The Hamiltonian is given by eq.~(\ref{eq:H0}) with $\lambda(\p,\sigma)$ and $\dual{\lambda}(\p,\sigma)$ replaced by their higher-spin generalizations. Here, the proportionality constants for $\lambda^{S,A}_{j}(\p,\sigma)$ are chosen such that the theory is local and respects the spin-statistics. Additionally,  the norms of the self-conjugate and anti-self-conjugate functions naturally admits the correct signatures such that the vacuum energy for bosons and fermions are positive and negative respectively.

\section{Conclusions} \label{con}

The theory of Elko and mass dimension one fermion was an unexpected discovery made by Ahluwalia and Grumiller in 2005. Since then, the theory has been extended beyond its original domain of particle physics to cosmology and mathematics.

In this review, we have provided a self-contained introduction on the kinematics of spin-half Elko, mass dimension one fermions and their higher-spin generalizations. At the kinematics level, the theory is physical in the sense that the free Hamiltonians are positive-definite and the fields commute/anti-commute with their adjoints at space-like separation. These properties provide a solid foundation towards formulating a consistent interacting theory.

The theory in its original formulation is Lorentz-violating. By introducing a new dual supplemented with the infinitesimal deformation, the theory becomes Lorentz-invariant~\cite{Ahluwalia:2016rwl}. In our opinion, while this represents important progress, it is not yet the complete solution. A complete solution should derive Elko from symmetry considerations following the formalism presented in ref.~\cite{Weinberg:1995mt}. Towards this end, we feel that it is also important to study the Lorentz-violating theory and identify the underlying space-time symmetries. The Lorentz-violation encountered here is encoded in the $\mathcal{G}(\phi)$ matrix which has many interesting properties in its own right. In particular, one should attempt a path-integral formulation based on eqs.~(\ref{eq:ff}) and (\ref{eq:fp}). This would  provide valuable insights to the existing theory.

An important problem is to formulate a consistent interacting theory. In ref.~\cite{Lee:2015sqj}, it was shown that using the standard $S$-matrix prescription, the Yukawa interaction involving mass dimension one fermions violates unitarity at one-loop. This problem can be resolved by introducing a new prescription to compute observables. The results will be presented elsewhere in future publications. 

The condition we have imposed on the Lagrangian $\mathscr{L}=\mathscr{L}^{\ddag}$ will also impose constraints on the allowed interactions. Further investigations on the properties of $\ddag$ is required. But based on what we know so far, it is possible to formally obtain certain conditions. Given a conserved matter current that satisfies $J_{\mu}(x)=J^{\ddag}_{\mu}(x)$, the vector field $A_{\mu}(x)$ which couples to the current must then satisfy $A^{\ddag}_{\mu}(x)=A_{\mu}(x)$. Depending on the definition of $\ddag$, this would constrain the possible interactions between mass dimension one particles and the SM.

The higher-spin construct is only one of the many possible generalizations. According to the Lounesto classification, there exists six different classes of spinors from which Elko belongs to the so-called flag-pole spinors. Following the works of Lounesto, one should systematically classify all higher-spin representations of the Lorentz group and study the corresponding quantum field theories. From what has been discovered so far, we expect the classification of higher-spin representations to also be endowed with rich structures and possibly provide a platform to explore physics beyond the SM. An even more expansive program is to study Elko and its generalizations in curved space-time. On this front, the gravitational dynamics of Elko and their applications to inflationary cosmology have been extensively studied in the literature though a systematic study remains to be carried out. In particular, it would be interesting to explore whether Elko furnishes representations associated with the maximally symmetric space-times in general relativity.
\appendix
\section{Non-deformed theory}\label{ndf}

The original theory without the deformed field adjoint is formulated using $\mathfrak{f}_{j}(x)$ and $\widetilde{\mathfrak{f}}_{j}(x)$ where
\begin{equation}
\widetilde{\mathfrak{f}}_{j}(x)=(2\pi)^{-3}\int\frac{d^{3}p}{\sqrt{2mE}}\sum_{\sigma}
\left[e^{-ip\cdot x}\widetilde{\lambda}^{S}_{j}a^{\ddag}(\p,\sigma)+e^{ip\cdot x}\widetilde{\lambda}^{A}_{j}(\p,\sigma)b(\p,\sigma)\right].
\end{equation}
Using eqs.~(\ref{eq:hs1}) and (\ref{eq:hs2}), we obtain
\begin{equation}
[\mathfrak{f}_{j}(t,\x),\widetilde{\mathfrak{f}}_{j}(t,\y)]_{\pm}=O.
\end{equation}
The propagator and Lagrangian are given by
\begin{eqnarray}
&&S_{j}(x-y)=\frac{i}{2}\int \frac{d^{4}q}{(2\pi)^{4}} e^{-iq\cdot(x-y)}\frac{I+\mathcal{G}_{j}(\phi)}{q^{2}-m^{2}+i\epsilon},\\
&&\mathscr{L}=\partial^{\mu}\widetilde{\mathfrak{f}}_{j}\partial_{\mu}\mathfrak{f}_{j}-m^{2}\widetilde{\mathfrak{f}}_{j}\mathfrak{f}_{j}
\end{eqnarray}
so that the conjugate momentum is $\mathfrak{p}_{j}(x)=\partial\widetilde{\mathfrak{f}}_{j}(x)/\partial t$. A straightforward calculation yields the following locality anti-commutator/commutators 
\begin{eqnarray}
&& [\mathfrak{f}_{j}(t,\x),\mathfrak{f}_{j}(t,\y)]_{\pm}=[\mathfrak{p}_{j}(t,\x),\mathfrak{p}_{j}(t,\y)]_{\pm}=O,\label{eq:ff}\\
&& [\mathfrak{f}_{j}(t,\x),\mathfrak{p}_{j}(t,\y)]_{\pm}=\frac{i}{2}\int \frac{d^{3}p}{(2\pi)^{3}}\, e^{i\mathbf{p\cdot(x-y)}}\left[I+\mathcal{G}_{j}(\phi)\right].\label{eq:fp}
\end{eqnarray}

%

\label{Bibliography}
\bibliographystyle{JHEP}  
\bibliography{Bibliography}  
\end{document}